\documentclass[10pt,a4paper]{article}

\usepackage{jheppub}

\usepackage{epsfig,epsf}
\usepackage{amsmath}
\usepackage{amsthm}
\usepackage{amsfonts}
\usepackage{amssymb}
\usepackage{dsfont}
\usepackage{epstopdf}
\usepackage{multirow}

\usepackage{marvosym}

\usepackage{slashed}

\usepackage[active]{srcltx}
\usepackage{hyperref,setspace}

\def\II{\hbox{{1}\kern-.25em\hbox{l}}}

\newcommand \vev [1] {\langle{#1}\rangle}
\newcommand \widebar [1] {\overline{#1}}
\def\II{\hbox{{1}\kern-.25em\hbox{l}}}
\DeclareMathOperator{\tr}{tr}

\title{{
\begin{flushright}
  \large \textnormal{\textrm{DESY 16-197}}\\[2mm]
  \large \textnormal{\textrm{LMU-ASC 51/16}}\\[5cm]
\end{flushright}
}
Higher-spin currents in the Gross-Neveu model at $1/n^2$}

\author[a,b]{A. N. Manashov,}
\author[c,d]{and E. D.  Skvortsov}

\affiliation[a]{
   Institut f\"ur Theoretische Physik, Universit\"at Hamburg,   D-22761 Hamburg, Germany}

\affiliation[b]{
   Institut f\"ur Theoretische Physik, Universit\"at
   Regensburg, D-93040 Regensburg, Germany}

\affiliation[c]{Arnold Sommerfeld Center for Theoretical Physics, Ludwig-Maximilians University Munich, \\
Theresienstr. 37, D-80333 Munich, Germany}

\affiliation[d]{Lebedev Institute of Physics,
Leninsky ave. 53, 119991 Moscow, Russia}

\emailAdd{alexander.manashov@desy.de}
\emailAdd{evgeny.skvortsov@physik.uni-muenchen.de}

\abstract{We calculate the anomalous dimensions of  higher-spin currents, both singlet and non-singlet,
in the Gross~-~Neveu model at the $1/n^2$ order.
It was conjectured that in the critical regime this model  is dual  to a higher-spin gauge
theory on $AdS_4$. The $AdS/CFT$ correspondence predicts that the masses  of  higher-spin fields
correspond to the scaling dimensions of the singlet currents in the Gross~-~Neveu model.}


\dedicated{\large In honour of Alexander A. Andrianov's 70th birthday}

\begin{document}

\maketitle

\section{Introduction}\label{Sec:Introduction}
Since its introduction in 1974~\cite{Gross:1974jv} the  Gross-Neveu (GN) model   serves as a toy model for studies
of many interesting physical phenomena. In particular, the GN model in $2<d<4$ dimensions possesses the
Wilson-Fisher fixed point. In the critical regime this model enjoys scale and conformal invariance and provides an
example of  nontrivial conformal field theory (CFT). Various critical indices in the GN model were calculated both
in the $2+\epsilon$ and
$1/N$ expansions (here $N$ is the number of the fermion field components), see e.g.
Refs.~\cite{Wetzel:1984nw,Gracey:1991vy,Luperini:1991sv,Kivel:1993wq,Derkachov:1993uw,Gracey:1992cp,Gracey:1990wi,Vasiliev:1992wr,Gracey:1993kc,Gracey:2008mf}.
Due to its simplicity, the GN model together with the $O(N)$ symmetric $\varphi^4$ model presents an ideal
playground for testing new techniques in
CFTs~\cite{El-Showk:2013nia,Rychkov:2015naa,Basu:2015gpa,Ghosh:2015opa,Raju:2015fza,Alday:2015ota,Skvortsov:2015pea,Giombi:2016hkj,Diab:2016spb,Hikida:2016wqj,Dey:2016zbg,Nii:2016lpa,Bashmakov:2016uqk,Gopakumar:2016wkt,Hikida:2016cla}.

In this work we compute the anomalous dimensions $\gamma_s$ of the specific composite operators, known as
higher-spin currents:
\begin{align}
J_s\equiv J_{\mu_1\ldots\mu_s}=\bar q^a \gamma_{\mu_1}\partial_{\mu_2}\ldots\partial_{\mu_s} q^a + \ldots.
\end{align}
Here $s$ is the spin and $q$ is the $N-$component Dirac fermion while the ellipses stand for the total derivative
terms and symmetrization over all indices and subtraction of traces are implied. In the free field approximation all the higher-spin currents are conserved and give rise to an infinite-dimensional symmetry, known as the higher-spin symmetry, which is
then broken by interactions, see e.g.
Refs.~\cite{Maldacena:2012sf,Rychkov:2015naa,Skvortsov:2015pea,Giombi:2016hkj}.

At the $1/N$ order the anomalous dimension of the current $J_s$ have been calculated in 1977 by Muta and Popov\'{i}c
\cite{Muta:1976js}. In this work we extend this calculation to the next order  using the technique developed
in~\cite{Vasiliev:1975mq,Vasiliev:1993ux,Derkachov:1997ch}. As a nontrivial check of our result we verify that the
structure of an asymptotic expansion of the anomalous dimension as a function of the conformal spin,
$j=s-1+(d+\gamma_s)/2$, agrees with the predictions of Refs.~\cite{Basso:2006nk,Alday:2015eya,Alday:2015ewa}.

Another surge of recent interest to the GN model comes from  studies of the AdS/CFT correspondence
\cite{Maldacena:1997re,Gubser:1998bc,Witten:1998qj}. It was conjectured in \cite{Klebanov:2002ja} that the critical
$O(N)$-vector model is dual to the higher-spin gauge theory in $\text{AdS}_4$. Shortly after, the conjecture was
extended to the GN model and its super-symmetric extensions \cite{Leigh:2003gk, Sezgin:2003pt}. Some non-trivial
tests of this conjecture were performed  at the level of three-point functions~\cite{Giombi:2009wh} and at the level
of one-loop determinants, see \cite{Giombi:2013fka,Giombi:2014yra} for a discussion and references. Remarkably, the
same $AdS_4$ higher-spin theory should be dual both to free and interacting models, depending on the boundary
conditions. Moreover, there is a continuous transition between fermionic and bosonic versions of these CFT's, i.e.
the three-dimensional bosonization \cite{Giombi:2011kc,Aharony:2011jz,Maldacena:2012sf}, which on the AdS side is
accounted for by a free parameter of the higher-spin theory and on the CFT side is realized via coupling to the
Chern-Simons sector \cite{Giombi:2011kc}. While the duality at the tree level is better supported, the most subtle
effects should come from the loops. According to \cite{Girardello:2002pp}, see also \cite{Ruhl:2004cf,Manvelyan:2008ks,Skvortsov:2015pea},
radiative corrections on the AdS side should generate masses $\delta m^2$ of the higher-spin fields that from the
CFT point of view correspond to the anomalous dimensions $\gamma_s$ of the higher-spin currents
\begin{align}\label{AdS-CFT}
m_s^2&=m_0^2(s)+\delta m_s^2,&
m_0^2(s)&=(d+s-2)(s-2)-s, &
\delta m_s^2=\gamma_s\Big(d-4+2s +\gamma_s\Big)\,,
\end{align}
The masses are measured in the units of the cosmological constant. Therefore, our results should be equivalent to two-loop computations in the higher-spin theory. More specifically,
one should be able to extract $\gamma_s$ from the logarithmic corrections to the near boundary behaviour of
higher-spin fields.

The paper is organized as follows: In Section~\ref{GN-model} we recall the definition of the GN model and briefly review the method to compute the critical exponents. The section~\ref{NLO-results} contains our results and
some details of calculations for the anomalous dimensions of the higher-spin currents at the next-to-leading order in
$1/N$. The renormalized propagators can be found in Appendix \ref{app:A}, while some of the numerical values of the
anomalous dimensions are collected in Appendix \ref{app:B}.

\section{GN model in $d$ dimensions}\label{GN-model}
The GN model with $U(N)$ symmetry describes a system of $d$ ($d\equiv 2\mu$)-dimensional $N$-component Dirac
fermions, $q\,(\bar q) \equiv \big\{q^a\,(\bar q^a), a= 1,...,N\big\}$. Its action (in Euclidean space) takes the
form~\footnote{The discussion of the issues related to  renormalization of this model in $2+\epsilon$ dimensions can
be found in Refs.~\cite{Gracey:2008mf,Vasiliev:1997sk,Vasiliev:1996rd,Vasiliev:1996nx,Gracey:2016mio}.}~\cite{Gross:1974jv}
\begin{align}\label{GNAd}
S &=-\int d^d x \left[\bar q\slashed{\partial}q + \frac {{g}}{2N} (\bar q q)^2\right ].
\end{align}
%
To generate a systematic $1/N$ expansion it is convenient to introduce  an auxiliary scalar field $\sigma$ and
rewrite GN action~\eqref{GNAd}  in the following form:
\begin{align}\label{N-GN}
S &=-\int d^d x\left[\bar q\slashed{\partial} q+\sigma \bar q q -\frac{N}{2g} \sigma^2\right]\,.
\end{align}
%
At a certain value of the coupling  $g=g_*$ the system undergoes the second order phase
transition~\cite{ZinnJustin:1991yn}. For $g < g_*$ the expectation value of $\sigma$ field vanishes,
$\sigma_0=\vev{\sigma}=0$,  and the fermions are massless, while for $g>g_*$  $\vev{\sigma}\neq 0$ and fermions
acquire mass, $m=\vev{\sigma}$ at the leading order. At the critical point, $g=g_*$, the correlators of the fields
$q,\bar q, \sigma$ exhibit  power law behaviour and, as it can be shown, the model enjoys  scale and conformal
invariance~\cite{Derkachov:1993uw}. Critical exponents are usually calculated with the help of the self-consistency
equations~\cite{Vasiliev:1981dg} or the conformal bootstrap~\cite{Vasiliev:1982dc} methods (see
Ref.~\cite{Vasilev:2004yr} for a review). However, it turns out that for the analysis of  the operators with nontrivial
tensor structure it is more convenient to use another approach described below.

In the infrared region (IR) (momenta much less than the cutoff $\Lambda$) the dominant contribution to the
propagator of the $\sigma$ field  in the leading order comes from the fermion loop~\cite{ZinnJustin:1991yn}
\begin{align}\label{sigma-prop}
D_\sigma(p)=-\frac1n {b(\mu)}/{(p^2)^{\mu-1}}\,,
&&
D_\sigma(x)=-\frac1n {B(\mu)}/{x^2}\,.
\end{align}
Here $n=N\times \tr \II$, where $\tr \II$ is  a trace of the unit matrix in the space of $d$-dimensional
spinors~\footnote{
 This trace does not have (and does not require) an exact expression in terms of $d$
 (for the integer dimensions $d=2,3,4$  it is usually assumed that  $\tr\II= 2,2,4$).
 }
 and the normalization factors are
\begin{align}\label{bmu}
 b(\mu)=(4\pi)^\mu\frac{\Gamma(2\mu-1)}{\Gamma^2(\mu)\Gamma(1-\mu)}\,, &&
B(\mu)=\frac{4\Gamma(2\mu-1)}{\Gamma^2(\mu)\Gamma(\mu-1)\Gamma(1-\mu)}\,.
\end{align}
For practical calculations   it is convenient to use  a simplified (massless) version of the GN model which is
critically equivalent to~\eqref{N-GN}. The action of the model is given by the following
expression~\cite{Vasiliev:1975mq,Vasilev:2004yr}
\begin{align}\label{L-GN}
S' &=-\int d^d x\left[\bar q\slashed{\partial} q -\frac12 \sigma L\sigma +\sigma \bar q q  + \frac12 \sigma L\sigma\right]\,.
\end{align}
The kernel $L$ is the inverse $\sigma$-propagator~\eqref{sigma-prop},
$L^{-1}= D_\sigma $. It has the form
\begin{align}
L(x)= \tr D_q(x) D_{q}(-x)=-n \left (\frac{\Gamma(\mu)}{2\pi^\mu}\right)^2 \frac1{(x^2)^{2\mu-1}}\,, &&
D_q(x) =-
\frac{\Gamma(\mu)}{2\pi^\mu}\frac{\slashed{x}}{[x^2]^{\mu}} \,,
\end{align}
where $D_q(x)$ is the fermion propagator.

The first two terms in~\eqref{L-GN} are considered as the free part of the action, $S_0$, and the remaining ones -- as an interaction,
$S_{int}$. The last term in~\eqref{L-GN} cancels diagrams with insertions of simple fermion loops in the
$\sigma$-lines. Of course, for giving a sense to the  model~\eqref{L-GN} it is necessary to introduce a regularization. Indeed, it
can be easily checked that the vertex $\sigma \bar q q $  diverges logarithmically  in any dimensions. A
regularization preserving the masslessness of propagators, that is important for practical calculations, was
proposed in~\cite{Vasiliev:1975mq}. Namely, in order to make diagrams finite it is sufficient to change the kernel $L$ in the
free part of the action, $S_0$, as follows
\begin{align}\label{Ldef}
L(x)\to L_\Delta(x)= L(x) (M^2 x^2)^{-\Delta} C^{-1}(\Delta)\sim x^{-2(2\mu-1+\Delta)}\,.
\end{align}
 Here $M$ is the scale parameter and $C(\Delta)$ is an arbitrary function regular at $\Delta=0$ such that
$C(0)=1$. The choice of the function $C(\Delta)$ affects only the normalization of  correlators but not their
scaling dimensions.

The UV divergences appear in diagrams as poles in $\Delta$ and are removed by the corresponding
counterterms. The renormalized action~\eqref{L-GN} takes the form
\begin{align}\label{L-GN-ren}
S'_{R} &=-\int d^d x\left[Z_1\bar q\slashed{\partial} q -\frac12 \sigma L_\Delta\sigma + Z_2 \sigma \bar q q
+ \frac12 \sigma L\sigma\right]\,.
\end{align}
The model  is not, however, multiplicatively  renormalized, i.e. $S'_R(q,\sigma)\neq S'(q_0,\sigma_0)$. This means
that the anomalous dimensions of the fields or composite operators are not related directly to the corresponding
renormalization factors. The multiplicative renormalizability can be restored in the extended model by introducing two
new charges~\cite{Vasiliev:1975mq},
\begin{align}\label{L-GN-ext}
S'_R(u,v)&=-\int d^d x\left[Z_1(u,v)\bar q\slashed{\partial} q -\frac u2 \sigma L_\Delta\sigma
+ Z_2(u,v) \sigma \bar q q  + \frac v2 \sigma L\sigma\right]\,,
\end{align}
so that $S'_R(q,\sigma,u,v) = S'(q_0,\sigma_0,u_0,v_0)$. Obviously, for $u=v=1$ the extended model coincides with the
model~\eqref{L-GN-ren}. Since the model~\eqref{L-GN-ext} is multiplicatively renormalizable the scale dependence of the Green functions
is described by  the renormalization group equations (RGEs). For instance,
let $\{\mathcal{O}_i\}$ be a set of operators which mix under renormalization.
The RGE for the $r$-point 1PI functions with the insertion of the
operators $\mathcal{O}_i$ takes the form
\begin{align}\label{RGE-I}
\Big( \big[ M\partial_M +\beta_u \partial_u+\beta_v\partial_v-n_\Phi \gamma_\Phi\big] \delta^{ik}+\gamma_{\mathcal{O}}^{ik}\Big)
\Gamma_k(u,v;p,p_1,\ldots,p_r)=0\,,
\end{align}
where  $n_\Phi=(n_q+n_{\bar q})\gamma_q+n_\sigma \gamma_\sigma$ and  the RG functions are defined in the standard  way
\begin{align}\label{RGf}
\beta_u=M\partial_M u\,, && \beta_v=M\partial_M v\,, && \gamma_\Phi=M\partial_M \ln Z_\Phi\,, &&
\gamma_{\mathcal{O}}=-M\partial_M \mathbf{Z} \mathbf{Z}^{-1}.
\end{align}
Here $Z_q=Z_1^{1/2}$, $Z_\sigma=Z_2 Z_1^{-1}$ and the matrix $\mathbf{Z}$ enters the definition of the renormalized
operator,
$\mathcal{O}_i^{R}(\Phi)=\mathbf{Z}_{ik} \mathcal{O}_k(\Phi_0)$. In an arbitrary subtraction scheme the term
\begin{align}\label{buv}
\big(\beta_u\partial_u+\beta_v\partial_v\big)\Gamma_i(u,v;\{p\})\Big |_{u=v=1}
=-2\gamma_\sigma (\partial_u+\partial_v)\Gamma_i(u,v;\{p\})\Big |_{u=v=1}\neq 0\,,
\end{align}
which implies that the RG functions $\gamma_\Phi,\gamma_{\mathcal{O}}$ are not  true anomalous dimensions that determine the scale
dependence of the correlators. Let us stress that the correlators in the model~\eqref{L-GN-ren} and in the
extended model
\eqref{L-GN-ext} at $u=v=1$  have  certain scaling dimensions, namely
\begin{align}
\Big( \big[ M\partial_M-n_\Phi \widebar \gamma_\Phi\big] \delta^{ik}+\widebar \gamma_{\mathcal{O}}^{ik}\Big)
\Gamma_k(u=v=1;\{p\})=0\,,
\end{align}
but, in general, $\widebar \gamma_\Phi \neq \gamma_\Phi$, $\widebar \gamma_{\mathcal{O}}\neq\gamma_{\mathcal{O}}$.
It was shown in~\cite{Vasiliev:1975mq} that in the MOM scheme the renormalized Green functions depend only on the
difference of the charges $u$ and $v$, $\Gamma_k(u,v;\{p\}))=\Gamma_k(u-v;\{p\})$. This implies that the term with
$\beta$-functions in the RGE~\eqref{RGE-I} vanishes and, 
hence, $\widebar \gamma_\Phi = \gamma^{MOM}_\Phi$, $\widebar \gamma_{\mathcal{O}} = \gamma^{MOM}_{\mathcal{O}}$.
Unfortunately  the calculations in the MOM scheme is hardly feasibly beyond the leading order. In the most suitable for
practical calculations  MS-scheme, ($Z-$factors are given by series  in $1/\Delta$)
in general, $\widebar \gamma_{\mathcal{O}}\neq\gamma_{\mathcal{O}}$. However as it was shown in~\cite{Derkachov:1997ch}
the difference is of order $1/n^3$
\begin{align}
\gamma_{\Phi}^*=\widebar \gamma_{\Phi}-\gamma^{\text{MS}}_{\Phi}=O(1/n^3)\,, &&
\gamma_{\mathcal{O}}^*=\widebar \gamma_{\mathcal{O}}-\gamma^{\text{MS}}_{\mathcal{O}}=O(1/n^3)\,.
\end{align}
Thus,  the anomalous dimensions in the MS scheme up to $1/n^2$ order inclusively can be calculated with the help of
Eqs.~\eqref{RGf}. Taking into account that
\begin{align}
\beta_u=2u \big(\Delta- \gamma_\sigma\big)\,, && \beta_v=-2v\gamma_\sigma
\end{align}
one derives (from now on we consider only the MS scheme and omit the label MS)
\begin{align}\label{gao}
\gamma_{\mathcal{O}}=-2\Big(\Delta u\partial_u-\gamma_\sigma(u\partial_u+v\partial_v)\Big) \mathbf{Z}\, \mathbf{Z}^{-1}
=-2u\partial_u \mathbf{Z}^{(1)}(u,v)|_{u=v=1}
\end{align}
where
\begin{align}
\mathbf{Z}=1+\sum_{k>0} {\Delta^{-k}} \mathbf{Z}^{(k)}\,.
\end{align}
Taking into account that there is no derivative with respect $v$ in \eqref{gao} one can put $v=1$ from the very
beginning~\footnote{ In this case, $v=1$, there are no diagrams with an insertion of the simple fermion loop into
the $\sigma$ lines (it is exactly cancelled by the term $1/2\sigma L\sigma$ in the action~\eqref{L-GN-ext}). }
arriving to the final expression for  anomalous dimensions~\cite{Derkachov:1997ch}
\begin{align}\label{gamma-O}
\gamma_{\mathcal{O}}=-2u\partial_u \mathbf{Z}^{(1)}(u,1)\Big|_{u=1}\,.
\end{align}
Taking into account that the charge $u$ appears only in the  $\sigma$ field propagator
\begin{align}
D_\sigma(x)=- \frac1{u}\times \frac1{n} B(\mu) C(\Delta) \frac{M^{2\Delta}}{(x^2)^{1-\Delta}}\,,
\end{align}
and
 $-u\partial_u$ counts a number of $\sigma$ lines in diagrams one concludes  that 
the contributions of each diagram to the $\mathbf{Z}$ factor and to the anomalous dimension, $\gamma_{\mathcal{O}}$,
differ by a factor
$2n_\sigma$, where $n_\sigma$ is the number of internal $\sigma$ lines in the diagram.

Equations~\eqref{gao},~\eqref{gamma-O} have a striking resemblance to the  analogous expressions
in the MS scheme in the dimensional regularization. Being a variation of  the standard RG technique, this
approach is rather effective for the analysis of composite operators with a nontrivial tensor structure. A more
detailed discussion can be found in
 Refs.~\cite{Derkachov:1997ch,Derkachov:1998js} and  a generalization to gauge theories in
 Refs.~\cite{Ciuchini:1999cv,Ciuchini:1999wy}.

On finishing the review we recall  known results for the anomalous dimensions of the fermion and auxiliary
fields~\cite{Gracey:1992cp,Gracey:1990wi,Vasiliev:1992wr,Gracey:1993kc}. We adopt the standard
notations~\cite{Vasilev:2004yr}
\begin{align}
\eta=2\gamma_q, && \gamma_\sigma=-\eta-\kappa\,.
\end{align}
The index $\eta$ is known with $1/n^3$ ~\cite{Vasiliev:1992wr,Gracey:1993kc} and $\kappa$ with
$1/n^2$~\cite{Gracey:1992cp,Vasiliev:1992wr} accuracy. We need the first two coefficients in  the expansion for
$\eta$, $\eta=\sum_{k\geq 1} \eta_k/n^k$,
\begin{align}
\eta_1 &=-{B(\mu)}/{2\mu}=-\frac{2\,\Gamma(2\mu-1)}{\Gamma(\mu+1)\Gamma(\mu)\Gamma(\mu-1)\Gamma(1-\mu)}\,,
\label{etaone}\\
\eta_2 &=\eta_1^2\, \frac12\frac1{(\mu-1)^2}\Big[\frac{(\mu-1)^2}{\mu}+3\mu+4(\mu-1)+2(\mu-1)(2\mu-1) \Psi(\mu)\Big]\,,\label{etatwo}
\end{align}
where $\Psi(\mu)=\psi(2\mu-1)-\psi(1)+\psi(2-\mu)-\psi(\mu)$,
and only the first one for $\kappa$:
\begin{align}
\kappa_1=\eta_1 \frac{\mu}{\mu-1}\,, && \gamma_\sigma^{(1)}=-\frac{2\mu-1}{\mu-1} \eta_1\,.
\end{align}
%

\section{Higher-spin currents}\label{NLO-results}
We consider the higher-spin (traceless and symmetric) operators bilinear in fermionic fields:
\begin{itemize}
\item the scalar (singlet) operators
\begin{align}
\mathcal{O}_s=\bar q \gamma_{\mu_1}\partial_{\mu_2}\ldots\partial_{\mu_s} q + \ldots.
\end{align}
\item the adjoint (non-singlet) operators
\begin{align}
\mathcal{O}_s^{A}=\bar q \, t^{A} \gamma_{\mu_1}\partial_{\mu_2}\ldots\partial_{\mu_s} q + \ldots.
\end{align}
\end{itemize}
Here $t^A$ are the generators of the $SU(N)$ group and  summation over isotopic indices is always implied, ($\bar
q\, q= \bar q^a\, q^a$). It is assumed that Lorentz indices are symmetrized and traces subtracted so that
$s$ is the spin of the operator.
The ellipses stand for the total derivatives which can be neglected if one is interested
in the anomalous dimensions only. The renormalized operators take the form
\begin{align}
[{\mathcal{O}}_s] ={\mathcal{Z}}_s \mathcal{O}_s\,, && [\mathcal{O}^A_s] = {\mathcal{Z}}_s^A \mathcal{O}_s^A\,.
\end{align}
Up to $1/n^2$ terms inclusively the anomalous dimensions are given by
\begin{align}
\gamma(s) =\eta  - 2u\partial_u\,\mathcal{Z}^{(1)}_s\Big|_{u=1}\,, && \gamma_A(s) =\eta  - 2u\partial_u\,\mathcal{Z}^{A,(1)}_s\Big|_{u=1}\,,
\end{align}
where  $\mathcal{Z}^{(1)}$ is a simple pole in the corresponding  renormalization factor $\mathcal{Z}=1+
\mathcal{Z}^{(1)}/\Delta+O(1/\Delta^2)$.

For odd $s$ the anomalous dimensions of singlet and non-singlet  operators coincide, $\gamma_s=\gamma_s^{A}$.
Therefore, from now on,  it will be tacitly implied that spin  $s$ is even for the singlet currents.

As usual, the renormalization factor $\mathcal{Z}$ is extracted from the correlator of the operator
with fermion fields at zero momentum transfer, $\vev{\mathcal{O}(0) q(p) \bar q(p)}$.
\begin{figure}[t]
\centerline{\includegraphics[width=0.70\linewidth]{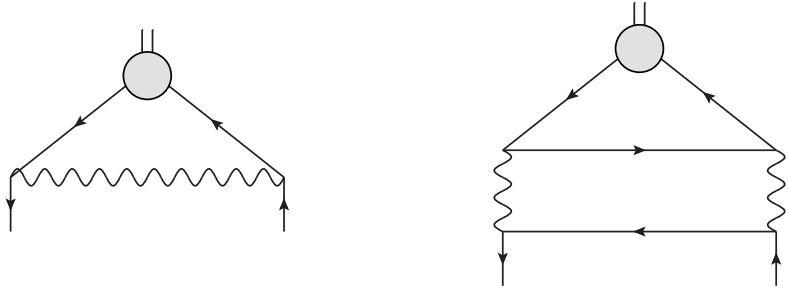}}
\caption{\sf The leading order diagrams for the correlator $\vev{\mathcal{O}(0) q(p) \bar q(p)}$.
The right diagram contributes only to the correlator of singlet operators of even spin.
}
\label{fig:LOD}
\end{figure}
Calculating  the leading order ($1/n$) diagrams  shown in Fig.~\ref{fig:LOD} we reproduce the result of Muta and Popovic~\cite{Muta:1976js}
\begin{subequations}
\label{MP}
\begin{align}
\gamma_{A}(s) & =\frac1n \eta_1\left( 1-\frac{\mu(\mu-1)}{(s+\mu-1)(s+\mu-2)}\right)+ O(1/n^2)\,,
\label{etaones}\\ \label{deltaetaones}
\gamma(s) &=\frac1n \eta_1
\left(1-\frac{\mu(\mu-1)}{(s+\mu-1)(s+\mu-2)}\left(1+\frac{\Gamma(2\mu-1)s!}{\Gamma(2\mu-3+s)(\mu-1)}\right)\right) + O(1/n^2)\,.
\end{align}
\end{subequations}
It can be easily checked that
the conserved currents, the spin-one  and spin-two singlet (the energy-momentum tensor) currents, have vanishing anomalous dimensions,
$\gamma_A(1)=0$ and $\gamma(2)=0$.

The anomalous dimensions (determined only for integer $s$) define analytic functions of complex variable (spin) $s$,
which
at integer points coincide with the corresponding anomalous dimensions. It is well known that such a continuation
should be done separately for even and odd $s$. Thus the Eqs.~\eqref{MP} define three analytic functions,
$\gamma_A^\pm (s)$ and $\gamma(s)$, where $\gamma^+_A, \gamma$ reproduce the corresponding anomalous dimensions  for even
$s$, and $\gamma^-_A(s)$ for odd $s$.  Obviously, in the $1/n$ order, $\gamma^+_A(s)=\gamma^-_A(s)$.

In  CFTs it is more natural  to consider  anomalous dimensions as  functions of the conformal spin $j$ defined as
\begin{align}\label{j-s-rel}
j=\frac12\big(\Delta_s+s\big)=\mu-1+s+\frac12\gamma(s)\,,
\end{align}
where $s$ is the spin and $\Delta_s$ is the scaling dimension of an operator.  For a given anomalous dimension
$\gamma(s)$ let us define  a function $f(j)$ as follows
\begin{align}
f(j)= f\left(\mu-1+s+\frac12\gamma(s)\right)=\gamma(s)\,.
\end{align}
It was noticed in~\cite{Basso:2006nk} that in all known examples  the large $j$ expansion of  the function $f(j)$
has a rather specific structure. 
Let us consider the $f$-functions  for the anomalous dimensions~\eqref{MP}. They take the form
\begin{align}
\gamma_A^\pm(s)=f^\pm_A(j) &=\frac1n \eta_1\left( 1-\frac{\mu(\mu-1)}{j(j-1)}\right)+ O(1/n^2)\,,
\notag\\
\gamma(s)=f(j) &=\frac1n \eta_1
\left(1-\frac{\mu(\mu-1)}{j(j-1)}\left(1+\frac{\Gamma(2\mu-1)}{\mu-1}\frac{\Gamma(j-\mu+2)}{\Gamma(j+\mu-2)}\right)\right) + O(1/n^2)\,.
\end{align}
At the leading order the functions $f_A^\pm$ are invariant under $j\to 1-j$. The singlet function $f(j)$ is given by the sum
of two terms one of which is invariant under $j\to 1-j$, but another  one, $\sim {\Gamma(j-\mu+2)}/{\Gamma(j+\mu-2)}$, is not.
The asymptotic expansion for this term has the form
\begin{align}\label{as-exp}
\left(j-\frac12\right)^{-2(\mu-1)}\sum_{k\geq 0} \frac{a_k}{(j(j-1))^k}\,.
\end{align}
Therefore, up to the prefactor the series is invariant under $j\to 1-j$. It was argued in
\cite{Alday:2015ewa,Alday:2015eya,Basso:2006nk} that a generic contribution to the asymptotic expansion of $f(j)$
has the structure~\eqref{as-exp} where  the coefficients $a_k$ are allowed to be a function
of $\ln (j-1/2)$.

Taking these findings into account we also present  our results for the anomalous dimensions as functions of conformal spin.
Besides that the corresponding expressions have a simpler form the very possibility to bring  results to the
form~\eqref{as-exp} provides  a nontrivial check of calculations.

\subsection{Non-singlet currents at $1/n^2$}
%
\begin{figure}[t]
\centerline{\includegraphics[width=0.67\linewidth]{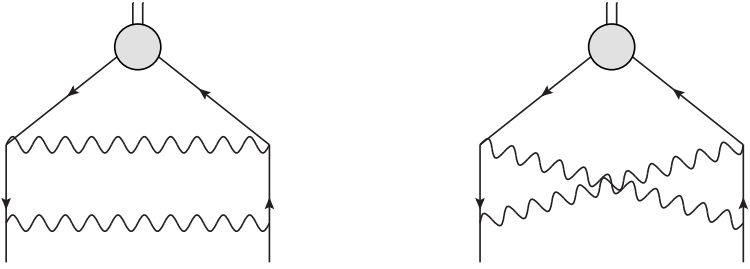}}
\caption{\sf The $1/n^2$ order diagrams for the non-singlet currents.
}
\label{fig:NLO-NS}
\end{figure}
Diagrams contributing
to the renormalization of the non-singlet current at $1/n^2$ order comprise
the diagrams  with the  self-energy (SE) and vertex
corrections to the leading order diagram (shown in the
l.h.s. of  Fig.~\ref{fig:LOD}) plus two additional diagrams
shown in Fig.~\ref{fig:NLO-NS}. The calculations are straightforward so that we present the answer only.
It is worth emphasizing that any diagram containing a fermion loop with odd number of attached $\sigma$
lines vanishes, which is exactly the feature that makes computations in the GN model more feasible that in the $O(N)$
$\sigma$-model.

\subsubsection{Anomalous dimensions}
The full conformal dimension of the non-singlet currents can be written as
\begin{align}
 \Delta_s &=\Delta_s^{(0)}+\frac1n\gamma^{(1)}(s)+\frac1{n^2}\gamma^{(2)}(s)+\ldots
 \notag\\
 &=
 2\mu+s-2+\frac{1}{n}\left(\eta_1+\gamma^{(1)}_s\right)+\frac{1}{n^2}\left(\eta_2+\gamma^{(2)}_s\right)
         +O\left(\frac1{n^3}\right)\,,
\end{align}
where $\eta_1$ and $\eta_2$ are given in Eqs.~\eqref{etaone} and \eqref{etatwo}, respectively, and  the anomalous dimension $\gamma_s^{(1)}$
takes the form
\begin{align}
    \gamma_s^{(1)}&=-\eta_1\frac{\mu  (\mu -1)}{(\mu +s-1) (\mu +s-2)}\,.
    \label{deltagammaone}
\end{align}
For the $1/n^2$ order anomalous dimension $\gamma^{(2)}_s$ we found
\begin{align}\label{anomnonsing}
    \gamma^{(2)}_s &= \gamma_s^{(1)}\eta_1\Big\{\psi(s+\mu -2)-\psi(\mu +1)+\frac{2 \mu -1}{\mu -1}
    \big[\psi(2 \mu -1)+\psi(-\mu )-\psi(\mu )-\psi(1)\big]
\notag
\\[1mm]
    &\quad -\frac{\mu(\mu -1)  }{2 (s+\mu -1) (s+\mu -2)}\left(1-\frac{1}{s+\mu -1}
    -\frac{1}{s+\mu -2}\right)+\frac{\mu }{(\mu -1) (s+\mu -2)} +\frac{1}{2}\frac{ \mu }{\mu -1}
\notag\\[2mm]
    &\quad-\frac{\mu }{\mu -1}\left[\psi(\mu )-\psi(s+\mu -2)\right]\Big\}
    -\frac{\eta_1^2}{2} \mu ^2  \left(1-\frac{(\mu -1)^2}{(s+\mu-1 ) (s+\mu -2)}\right) R_s(\mu)\,,
\end{align}
where $R_s(\mu)$ is (cf. with Eq.~(5.11) in ~\cite{Derkachov:1997ch})
\begin{align}\label{Rsmu}
 R_s(\mu)&=\int_0^1 d\alpha\, (1-\alpha)^{\mu-2}\int_0^1 d\beta\, (1-\beta)^{\mu-2}(1-\alpha-\beta)^{s-1}\,.
\end{align}
The last term in (\ref{anomnonsing}), the only contribution  which can not be expressed in terms of Euler's $\psi$-function,
comes entirely from the diagram in the r.h.s. in Fig~\ref{fig:NLO-NS}.
Finally, as a simple consistency check one can verify that $\gamma^{(2)}(s=1)\equiv \gamma_{s=1}^{(2)}+\eta_2=0$.

\begin{figure}[t]
\centerline{\includegraphics[width=0.87\linewidth]{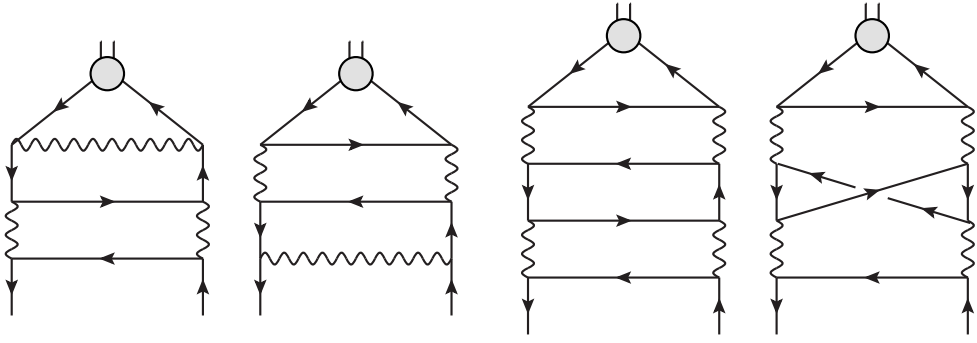}}
\caption{\sf The $1/n^2$ order diagrams for the singlet currents.
}
\label{fig:NLO-SS}
\end{figure}
%

\subsection{Singlet currents at $1/n^2$}

In order to find the anomalous dimensions  of the singlet currents at $1/n^2$ order one has to calculate diagrams shown in
Fig.~\ref{fig:NLO-SS} and the diagrams with all  possible self-energy and vertex
insertions to   the rightmost diagram in Fig.~\ref{fig:LOD}. The calculation  does not bring about
any troubles and can be easily performed with the help of the standard methods, see Ref.~\cite{Vasilev:2004yr} for a review.

\subsubsection{Anomalous dimensions}
The anomalous dimensions of the singlet currents with odd spins are equal to those of non-singlet ones.
The scaling dimensions of the currents with even spins can be written as
\begin{align}
 \Delta_s&=2\mu+s-2+\frac{1}{n}\left(\eta_1+\gamma^{(1)}_s+\Delta\gamma^{(1)}_s\right)
 +\frac{1}{n^2}\left(\eta_2+\gamma^{(2)}_s+\Delta \gamma^{(2)}_s\right)+O\left(\frac1{n^3}\right)\,.
\end{align}
The indices $\eta_{1,2}$ are defined in Eqs.~\eqref{etaone}, \eqref{etatwo},  $\gamma_s^{(1)}$,
$\gamma^{(2)}_s$ -- in Eqs.~\eqref{deltagammaone}, \eqref{anomnonsing}, respectively,  and $\Delta\gamma_s^{(1)}$ is the additional shift in
 \eqref{deltaetaones} for the singlet currents at order $1/n$:
\begin{align}
 \Delta\gamma^{(1)}_{s}&=-\eta_1\frac{\mu  \Gamma (2 \mu -1) \Gamma (s+1)}{(\mu +s-2) (\mu +s-1) \Gamma (s+2 \mu -3)}\,.
\end{align}
For the second order correction $\Delta\gamma^{(2)}_{s}$ 
we found
{\allowdisplaybreaks
\begin{align}\label{ansinglet}
    \Delta\gamma^{(2)}_s&=\eta_1\Delta\gamma_s^{(1)} \Biggl\{ 2\frac{2 \mu -1}{\mu -1}
\Big(\psi(2\mu-1)+\psi(-\mu)-\psi(\mu)-\psi(1)\Big)
\notag\\
&\quad
-\frac12\biggl[
3\Big(\psi(2\mu+s-3)-\psi(\mu+s-2)+\psi(2-\mu)-\psi(2)\Big)+
\psi(s)-\psi(1)
+\frac{1}{s+2\mu-3}
\notag\\
&\quad
+\psi(\mu+s-2)-\psi(\mu)
+4\frac{2\mu-1}{\mu(\mu-1)}+1
\biggr]
\notag\\
&\quad
-\frac{\mu}{\mu-1} \biggl[ \psi(s) 
+\psi(2\mu-2+s)-2\psi(s+\mu-1)+\psi(2-\mu)+\psi(\mu)-2\psi(1)
\biggr]\Biggr\}
\\
&\quad
-\frac12\gamma_s^{(1)}\Delta \gamma_s^{(1)}\biggl\{ \frac{\mu+s-2}{(\mu-1)(2\mu+s-3)}
+2\left(-1+\frac1{\mu+s-1}+\frac1{\mu+s-2}\right)
\notag\\
&\quad
+\psi(2\mu-3+s)-\psi(\mu)+ 
\psi(1-\mu)-\psi(s)
\biggr\}
\notag\\
&\quad
-\frac12(\Delta \gamma_s^{(1)})^2\biggl\{ \frac{1}{s(2\mu+s-3)}
+\left(-1+\frac1{\mu+s-1}+\frac1{\mu+s-2}\right)
\notag\\[2mm]
&\quad
+\psi(2\mu+s-3)-\psi(\mu)+\psi(2-\mu)-\psi(s+1)
\notag\\[2mm]
    &
    -\frac{(\mu+s-1)(\mu+s-2)}{s(\mu-1)(s+2\mu-3)}
    \Big[\psi(s+2\mu-3)+\psi(s+1)-2\psi(s+\mu-1)+\psi(\mu)-\psi(1)-J_s(\mu)
\Big]\Biggr\},
\notag
\end{align}}
where the function $J_s(\mu)$ is defined as
\begin{align}\label{Jsmu}
 J_s(\mu)&=\frac{\Gamma(s+\mu-1)}{s!\Gamma(\mu-2)} \int_0^1d\alpha \, \alpha^{2\mu-4+s} \int_0^1d\beta\,
 \frac{\beta^{\mu-2}(1-\beta)^{s}}{1-\alpha \beta}.
\end{align}
Again, the contribution in the last line in \eqref{ansinglet}, which can not be expressed in terms of $\psi$ functions only, is due
the rightmost diagram in Fig.~\ref{fig:NLO-SS}.
Finally, taking into account that
\begin{align}
  J_2(\mu)&=
  \frac12{ \left(-2 S_1({\mu -2})+2 S_1({2 \mu -2})+\frac{1+\mu }{1-\mu }\right)}\,.
\end{align}
one can check that the stress-tensor is conserved, i.e. $\eta_2+\gamma^{(2)}_{s=2}+\Delta\gamma^{(2)}_{s=2}=0$.

\subsection{Conformal spin expansion}
\label{sec:confspin}
Below we check that the anomalous dimensions of singlet and non-singlet currents can be cast into the form \eqref{as-exp} when expressed as functions of conformal spin.

\subsubsection{Conformal spin expansion of non-singlet currents}
In order to express the anomalous dimensions in terms of conformal spin we write
\begin{align}\label{NLO-NSO}
\gamma^\pm_A(s)=f^\pm_A(j)=\eta \Big(1 + f_1(j)\Big) + \eta^2 f_2^\pm(j)\,,
\end{align}
where $\eta=\eta_1/n+\eta_2/n^2+\ldots$ and $j$ is defined in Eq.~\eqref{j-s-rel}. Note that the first term in~\eqref{NLO-NSO} contains both $1/n$ and  $1/n^2$ terms.
For the functions $f_1$ and $f_2^\pm$ we get
\begin{align}\label{NS1}
f_1(j) & =-\frac{\mu(\mu-1)}{j(j-1)}\,,
\\[2mm]
\label{NS2}
f_2^\pm(j) &= \frac12 
\frac{\mu^2-\mu+1}{\mu(\mu-1)} f^2_1(j)
+f_1(j)\Biggl\{
\frac{2\mu-1}{\mu-1}
\Big(\psi(j)-\psi(\mu)\Big)
+\frac{\mu-1}{2\mu}-\frac1{2(\mu-1)^2}
\Biggr\}
\notag\\[2mm]
&\quad
-\frac12\mu^2 \left[1-\frac{(\mu-1)^2}{j(j-1)}\right] \Big(R^+(j,\mu)\mp R^-(j,\mu)\Big)\,,
\end{align}
where
\begin{align}
R^-(j,\mu)= {\Gamma^2(\mu-1)}\frac{\Gamma(j+1-\mu)}{\Gamma(j+\mu-1)}\,,
&&
R^+(j,\mu)  
=\frac1{j} \int_0^1d u \,  \bar u^{j-2} \,  {}_2F_1 \genfrac{(}{|}{0pt}{}{2-\mu,1}{j+1}-\frac{u}{\bar u}\biggl)\,.
\end{align}
The functions $R^\pm(j,\mu)$ are related to the function $R_s(\mu)$ as follows:
\begin{align}\label{R-integer}
R^+(j_s,\mu)+(-1)^{s-1} R^-(j_s,\mu)=R_s(\mu)\,,
\end{align}
where $j_s=s+\mu-1$. Note also, that in this formulation  the spin-one current conservation is equivalent to the
constraint  $f^-_2(\mu)=0$.

One can verify that the large $j$ behavior of anomalous dimensions \eqref{NLO-NSO} agrees with the predictions of
Refs.~\cite{Alday:2015ewa,Alday:2015eya,Basso:2006nk}. It is easy to see for all terms except, may be, $R^+(j,\mu)$.
For this function  one can use the Mellin - Barnes representation for hypergeometric function to get asymptotic expansion
\begin{align}
R^+(j,\mu) 
&=
\frac1{2\pi i} \int_{-i\infty}^{i\infty} dt
\frac{\Gamma(2-\mu+t)\Gamma^2(1+t)\Gamma(-t)\Gamma(j-t-1)}{\Gamma(2-\mu)\Gamma(j+t+1)}
\underset{j\to\infty}{\simeq}\sum_{k\geq 0} c_k \, 
\frac{\Gamma(j-k-1)}{\Gamma(j+k+1)}\,,
\end{align}
with $c_k=(-1)^k{k!\Gamma(2-\mu+k)}/{\Gamma(2-\mu)}$.

\subsubsection{Conformal spin expansion of singlet currents}
We can rewrite the answer for the singlet current in the form
\begin{align}\label{NLO-SO}
\gamma(s)=f(j)=\eta \Big(1 + f_1(j)+\Delta f_1(j)\Big) + \eta^2\Big(f^+_2(j) +\Delta f_2(j)\Big) +O(1/n^3)\,,
\end{align}
where $j=s+\mu-1+\gamma(s)/2$, the  functions $f_1(j),\, f_2^+(j)$ are defined by Eqs.~\eqref{NS1}, \eqref{NS2} and
\begin{align}
\Delta f_1(j) & =  f_1(j)\,\frac{\Gamma(2\mu-1)}{\mu-1}\frac{\Gamma(j-\mu+2)}{\Gamma(j+\mu-2)}\,.
\end{align}
For the function $\Delta f_2(j)$ we obtained
\begin{align}
\Delta f_2(j) & =\Delta f_1(j)\Biggl\{-\Big[\psi(j)-\psi(\mu)\Big]
-\frac{2\mu-1}{\mu-1}\Big(\Psi(j,\mu)-\Psi(\mu,\mu)\Big)-\frac12\Big(\psi(2-\mu)-\psi(\mu)\Big)
\notag\\[1mm]
&\quad
-\frac12\frac1{j(j-1)}  +\frac{(2\mu-3)(3\mu-1)}{2(\mu-1)(j-\mu+1)(j+\mu-2) }
+\frac1{(\mu-1)^2}+\frac1{2\mu(\mu-1)}-1
\notag\\[2mm]
&\quad
-\frac12 f_1(j)
\biggl(\psi(1-\mu)-\psi(\mu-1)-2+
\frac{2\mu-3}{(j-\mu+1)(j+\mu-2) } 
\biggr)
\notag\\[1mm]
&\quad
-\frac12\Delta f_1(j)\Biggl( \psi(2-\mu)-\psi(\mu)-1+\frac{1}{(j-\mu+1)(j+\mu-2)}
\notag\\[1mm]
&\quad
-\frac{j(j-1)}{(\mu-1)(j-\mu+1)(j-2+\mu)} 
\Big[ \Psi(j,\mu)
+\psi(\mu)-\psi(1)
 - \mathrm{J}(j,\mu)\Big] \Biggr)
\Biggr\}\,,
\end{align}
where
\begin{align}
\Psi(j,\mu)=\psi(j-2+\mu)+\psi(j+2-\mu)-2\psi(j)
\end{align}
and $\mathrm{J}(j,\mu)$ is an analytic continuation of the function  $J_s(\mu)$, defined in~\eqref{Jsmu}, to non-integer spins:
$\mathrm{J}(j_s,\mu)=J_s(\mu)$, for $j_s=s+\mu-1$:
\begin{align}\label{MB2}
\mathrm{J}(j,\mu)&=
\frac{\Gamma(j)}{\Gamma(\mu-2)\Gamma(j+2-\mu)(j+\mu-2)}
\int_0^1 du u^{\mu-2} \bar u^{j-\mu} {}_2F_1 \genfrac{(}{|}{0pt}{}{1,1}{j+\mu-1} -\frac{u}{\bar u} \biggl)
\notag\\[2mm]
&=
\frac1{\Gamma(\mu-2)}\frac{\Gamma(j-2+\mu)}{\Gamma(j+2-\mu)}\frac1{2\pi i}
\int_{-i\infty}^{i\infty} dt\, {\Gamma^2(t+1)\Gamma(\mu+t)\Gamma(-t)}\frac{\Gamma(j-\mu+1-t)}{\Gamma(j+\mu-1+t)}\,.
\end{align}
It can be checked that $f^+_2(j) +\Delta f_2(j)$ vanishes for $j=\mu+1$, so that the anomalous dimension of the
energy-momentum  tensor, $\gamma(s=2)=0$, as it should be. Next, taking into
account that for large $j$
$$
\Psi(j,\mu)-\Psi(1-j,\mu) = O(e^{-\pi|\text{Im} j|})
$$
 and, as it follows from Eq.~\eqref{MB2},
\begin{align}
\mathrm{J}(j,\mu)\underset{j\to\infty}{\simeq} \sum_{n\geq 0} c_n(\mu) \frac{\Gamma(j-2+\mu)}{\Gamma(j+2-\mu)}\frac{\Gamma(j-\mu+1-n)}{\Gamma(j+\mu-1+n)}
\end{align}
we conclude that the asymptotic expansion of the function $f(j)$ for large $j$ agrees with the predictions of
Ref.~\cite{Alday:2015ewa}.
\vskip 5mm
\vskip 3mm

\subsection{Three dimensions and higher-spin masses}
The case of three dimensions is of the most interest.
First of all, we give the expression for the index $\eta$ in $d=3$ model
\begin{align}
\eta_1=\frac{8}{3\pi^2}\,,&& \eta=\frac{\eta_1}n\left(1+\frac{28}{3n}\eta_1+\ldots\right).
\end{align}
The full order $1/n^2$ anomalous dimension of the non-singlet currents $\gamma^{(2)}_s$ can be simplified to
\begin{align}\label{nonsingAA}
\begin{aligned}
 \gamma^{(2)}_s&= 
 \frac{3\eta^2_1}{4 \left(4 s^2-1\right)}\Biggl\{\frac{3 \pi  (-1)^s \left(2 s^2-1\right)}{s}
 -\frac{2 (24 s+9)}{4 s^2-1}-\frac{48 s}{\left(4 s^2-1\right)^2}+2\\
 &-32 \log (2)+\frac{3 \left(2 s^2-1\right)}{s} {\biggl[S_1\left( \tfrac{s}{2}-\tfrac{3}{4}\right)
 -S_1 \left(\tfrac{s}{2}-\tfrac{1}{4}\right)\biggr]}-16 S_1\left({s-\tfrac{3}{2}}\right)\Biggr\}\,,
\end{aligned}
\end{align}
where $S_1(j)\equiv \psi(j+1)-\psi(1)$.
The singlet anomalous dimensions can also be simplified:
\begin{align}\label{singAA}
\begin{aligned}
 \Delta\gamma_s^{(2)}&=\frac{3\eta_1^2}{4s^2-1}\Biggl\{-\frac{3 \left(144 s^3+92 s^2-28 s-15\right)}{2 (4 s^2-1)^2}+\frac{11}{2}-8 s-2 (14 s+3) \log (2)\\
 &-2 (7 s+3) S_1\left({s-\tfrac{3}{2}}\right)+2(8 s+3) S_1(s-1)+3 \biggl[S_1\left({\tfrac{s-1}{2}}\right)-S_1\left({\tfrac{s-2}{2}}\right)\biggr]\Biggr\}\,.
 \end{aligned}
\end{align}
The conservation of the stress-tensor corresponds to $\eta_2+\gamma^{(2)}_{s=2}+\Delta\gamma_{s=2}^{(2)}=0$.

\begin{figure}[t]
\centerline{\includegraphics[width=0.45\linewidth]{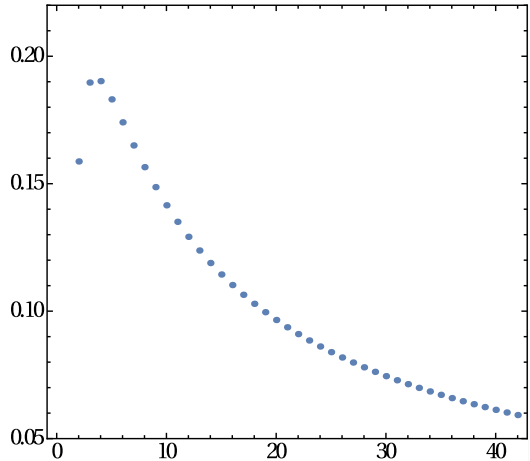}}
\caption{\sf The function $\varkappa_1(2k)$,~{Eq}.~\eqref{varkappa1}.
}
\label{fig:kappa}
\end{figure}

Using the above results for the anomalous dimensions of the currents in the critical $d=3$ GN model
one can  derive   masses of the higher-spin gauge fields~\eqref{AdS-CFT} in the dual $AdS_4$ model.
Plugging the first order anomalous dimensions  and \eqref{nonsingAA}, \eqref{singAA} into \eqref{AdS-CFT}
one gets
\begin{align}
\begin{aligned}
\delta m_s^2&=\frac{2}{n}\eta_1(s-2)+\frac{\eta_1^2}{n^2}\frac{1}{s(1+2s)}\Biggl\{\frac{9}{4} \pi  \left(2 s^2-1\right)+\frac{s \left(224 s^3-244 s^2+88 s-317\right)}{3 (2 s-1)}+\\&+\frac{9}{4} \left(2 s^2-1\right)\left[ S_1{\left(\tfrac{s}{2}-\tfrac{3}{4}\right)}- S_1{\left(\tfrac{s}{2}-\tfrac{1}{4}\right)}\right]+9 s\left[ S_1{\left(\tfrac{s-1}{2}\right)}- S_1{\left(\tfrac{s-2}{2}\right)}\right]+\\
&+6 s (8 s+3) S_1{\left(s-1\right)}-6 s (7 s+5) S_1{\left(s-\tfrac{3}{2}\right)}-42 s (2 s+1) \log (2)\Biggr\}\,.
\end{aligned}
\end{align}
One can see that the graviton remains massless, $\delta m_{s=2}^2=0$, as it should be.
For large spin 
the mass-spin dependence for the higher-spin fields,  Eq.~\eqref{AdS-CFT}, can be written in the form
\begin{align}
\delta m^2_s &= 2\eta (s-2) \Big(1+\eta \varkappa_1(s)+\ldots\Big)\,.
\end{align}
Thus  in the leading order this dependence  takes the form of linear Regge trajectory
 with the slope $\alpha^\prime=1/2\eta$. Note that the same linear mass squared spin dependence holds also in the $O(N)$
 model~\cite{Ruhl:2004cf}.
The deviation from the linear trajectory, $\varkappa_1(s)$, is due to the next-to-leading corrections,
\begin{align}\label{varkappa1}
\varkappa_1(s) & =\frac1{2s-1}+\frac{2s-1}{2(s-2)}\Big(f_2^+(s+1/2)+\Delta f_2 (s+1/2)\Big)\,.
\end{align}
The correction $\varkappa_1(s)$ is positive for even spins, see Fig.~\ref{fig:kappa}, and
vanishes as
$\varkappa_1(s) = \frac32\ln s/s +\ldots $ for  large spin.

\section{Summary}
We have calculated the $1/n^2$ corrections to the scaling dimensions of the (non)singlet higher-spin currents in the GN
model. As nontrivial checks we found that the spin-one non-singlet current and the spin-two stress-tensor current are conserved
and the asymptotic expansion in terms of conformal spin agrees with the  results of Refs.~\cite{Basso:2006nk,Alday:2015eya,Alday:2015ewa}.

In three dimensions the anomalous dimensions can be considerably simplified. Some of the numerical values can
be found in Appendix \ref{app:B}, which should facilitate comparison with other methods,
e.g. the numerical bootstrap along the lines of \cite{ElShowk:2012ht,Iliesiu:2015qra,El-Showk:2014dwa}.

Given the anomalous dimensions of the singlet currents we also computed the loop corrections to
the masses of higher-spin fields in the four-dimensional higher-spin theory dual to the GN model, known as Type-B.
At the leading order the mass spin dependence  has the form of a linear Regge  trajectory while the next-to-leading
correction gives rise to deviation from linearity.

As was observed in \cite{Giombi:2011ya,Bekaert:2012ux}, there is some diagrammatic dictionary between all the Feynman-Witten
graphs at order $1/N^k$ in the bulk and Feynman graphs at the same order on the CFT side.
We note that  the graphs of certain topologies present in the bosonic $O(N)$-model are absent in the Gross-Neveu model
since they contain traces of odd number of $\gamma$-matrices. This fact indicates that in $d>3$ the Type-B higher-spin theory should enjoy some hidden
simplicity as compared to the type-A case, which is dual to the bosonic $O(N)$ model. Also, as it is clear already from the order $1/n$ results, there are several
contributions to the anomalous dimensions
which differ by their large spin asymptotic. It would be interesting to understand this effect from the bulk side.

\section*{Acknowledgments}
\addcontentsline{toc}{section}{Acknowledgments}

The work of A.M. was supported in part by Deutsche Forschungsgemeinschaft (DFG) with the grant MO~1801/1-1.
The work of E.S. was supported  in part  by the Russian Science Foundation grant 14-42-00047 in association with Lebedev Physical Institute and by the DFG Transregional Collaborative Research Centre TRR 33 and the DFG cluster of excellence ”Origin and Structure of the Universe”. E.S. would like to thank Munich Institute for Astro- and Particle Physics (MIAPP) of the DFG cluster of excellence "Origin and Structure of the Universe" for the hospitality. E.S. is also grateful to Simone Giombi and Volodya Kirilin for useful discussions of (Chern-Simons) vector-models. 

\begin{appendix}
\renewcommand{\thesection}{\Alph{section}}
\renewcommand{\theequation}{\Alph{section}.\arabic{equation}}
\setcounter{equation}{0}\setcounter{section}{0}

\section{Renormalized Propagators }\label{app:A}
For completeness of exposition we give here expressions for the renormalized propagators in the $1/n$ order
assuming that the normalization factor $C(\Delta)$ is chosen $C(\Delta)=1$.
The propagators take the form
\begin{align}
D_q^{-1}(p)  = i\slashed{p} \left({p^2}/{\widebar M^2}\right)^{-\gamma_q}\, A_q(\mu)\,,
&&
D^{-1}_\sigma(p) & = -n b^{-1}(\mu) p^{2(\mu-1)} \left({p^2}/{\widebar M^2}\right)^{-\gamma_\sigma}\, A_\sigma(\mu)\,,
\end{align}
where $\widebar M=2M$, $b(\mu)$ is defined by Eq.~\eqref{bmu} and
\begin{align}
A_q(\mu) & = 1+\gamma_q/\mu(\mu-1)+O(1/n^2)\,,
\notag\\
A_\sigma(\mu) & =1+\gamma_{\sigma} \Big(\psi(2\mu-1)+\psi(-\mu)-\psi(\mu)-\psi(1)\Big)+O(1/n^2)\,.
\end{align}

\section{Numerical Values}\label{app:B}
Since the formulas for the anomalous dimensions are quite complicated we list below few numerical values in one of
the most interesting cases of three dimensions.
The order $1/n$ results are due to \cite{Muta:1976js} and we collect the order $1/n^2$ coefficients only.
It is worth stressing that we give below the full anomalous dimensions at order $1/n^2$, i.e. $\eta_2$ is included.
It is convenient to measure the anomalous dimensions in the units of $\eta_1^2=64/(9\pi^4)$.
The spin-one current is always conserved, $\gamma^2_A(1)=0$, and for few other we find
\begin{align}
  \gamma^{(2)}_A(2)&=\frac{1104}{125}\approx8.832 &
 \gamma^{(2)}_A(3)&=\frac{912896}{128625}\approx 7.09734 \\
 \gamma^{(2)}_A(4)&=\frac{1324432}{138915}\approx9.53412&
 \gamma^{(2)}_A(5)&=\frac{154300672}{18866925}\approx 8.17837
\end{align}
The anomalous dimensions of the singlet currents with odd spins are the same as for the non-singlet ones.
Below are the full anomalous dimensions at order $1/n^2$ for even spins currents. Stress-tensor is conserved,
i.e. $\gamma_S^2(2)=0$, and for a few others we have in the units of $\eta_1^2$
\begin{align}
 \gamma_{S}^{(2)}(4)&=\frac{16600}{3087}\approx5.37739 & \gamma_{S}^{(2)}(6)&=\frac{12495584}{1816815}\approx6.87774 \\
 \gamma_{S}^{(2)}(8)&=\frac{145039504}{19144125}\approx7.57619 & \gamma_{S}^{(2)}(10)&=\frac{133304287652}{16712124975}\approx7.9765
\end{align}

\end{appendix}
\setstretch{1.0}
\bibliographystyle{utphys}
\bibliography{megabib.bib}

\end{document}